\documentclass[]{pasj02} 
\usepackage[switch,mathlines]{lineno} % add line number to manuscript

\jyear{2025}
\Received{2025/08/13}%{yyyy/mm/dd}
\Accepted{2025/11/20}%{yyyy/mm/dd}
\begin{document} 

\title{Peculiar SN Ic 2022esa: An explosion of a massive Wolf-Rayet star in a binary as a precursor to a BH-BH binary?\thanks{This research is based in part on data collected at the Subaru Telescope, which is operated by the National Astronomical Observatory of Japan. We are honored and grateful for the opportunity of observing the Universe from Maunakea, which has the cultural, historical, and natural significance in Hawaii.}}

%%% begin:list of authors
% Do NOT capitalize all letters in "textsc".
\author{
Keiichi \textsc{Maeda},\altaffilmark{1}\altemailmark\orcid{0000-0003-2611-7269} \email{keiichi.maeda@kusastro.kyoto-u.ac.jp} 
Hanindyo \textsc{Kuncarayakti},\altaffilmark{2,3}\orcid{0000-0002-1132-1366}
Takashi \textsc{Nagao},\altaffilmark{2,4,5}\orcid{0000-0002-3933-7861}
Miho \textsc{Kawabata},\altaffilmark{6}\orcid{0000-0002-4540-4928}
Kenta \textsc{Taguchi},\altaffilmark{1,6}\orcid{0000-0002-8482-8993}
Kohki \textsc{Uno},\altaffilmark{1}\orcid{0000-0002-6765-8988}
Kishalay \textsc{De}\altaffilmark{7,8}\orcid{0000-0002-8989-0542}
}

\altaffiltext{1}{Department of Astronomy, Kyoto University, Kitashirakawa-Oiwake-cho, Sakyo-ku, Kyoto, 606-8502, Japan}
\altaffiltext{2}{Tuorla Observatory, Department of Physics and Astronomy, University of Turku, FI-20014 Turku, Finland}
\altaffiltext{3}{Finnish Centre for Astronomy with ESO (FINCA), University
of Turku, FI-20014 Turku, Finland}
\altaffiltext{4}{Aalto University Mets\"ahovi Radio Observatory, Mets\"ahovintie 114, 02540 Kylm\"al\"a, Finland}
\altaffiltext{5}{Aalto University Department of Electronics and Nanoengineering, P.O. BOX 15500, FI-00076 AALTO, Finland}
\altaffiltext{6}{Okayama Observatory, Kyoto University, 3037-5 Honjo, Kamogatacho, Asakuchi, Okayama 719-0232, Japan}
\altaffiltext{7}{Department of Astronomy, Columbia University, Mail Code 5246, 538 West 120th Street, New York, NY 10027, USA}
\altaffiltext{8}{The Center for Computational Astrophysics, Flatiron Institute, 162 5th Ave., New York, NY 10010, USA}

%\footnotetext[$\dag$]{Present address: ....}

%%% end:list of authors

%% !!! Select 3 to 5 words from PASJ's key words !!! 
%% List of Key Words: https://academic.oup.com/pasj/pages/Pasj_Keywords 
%% "\KeyWords{ }" always has to be placed before ``\maketitle'' 

\maketitle

\begin{abstract}
A class of supernovae (SNe) termed `SN Ic-CSM' are characterized by late-time emergence of narrow emission lines of elements formed in the oxygen core of a massive star. A popular scenario is the interaction of the SN ejecta and O-rich circumstellar medium (CSM), i.e., Circumstellar Interaction (CSI). Uncovering the progenitor system of SNe Ic-CSM plays a critical role in understanding the final evolution of a massive star to a bare C+O star. In this Letter, we present observations of SN 2022esa which we show is an SN Ic-CSM. Surprisingly, a stable periodicity of $\sim 32$ days is found in its light-curve evolution with a hint of a slowly increasing period over $\sim 200$ days. We argue that the main power source is likely the interaction of the SN ejecta and O-rich CSM, while the energy input by the post-SN eccentric binary interaction within the SN ejecta is another possibility. In either case, we propose a massive Wolf-Rayet (WR) star as the progenitor, in a WR-WR or WR-BH (black hole) binary that will eventually evolve to a BH-BH binary. Specifically, in the CSI scenario, the progenitor system is an eccentric binary system with an orbital period of about a year, leading to the observed periodicity through the modulation in the CSM density structure. We also show that some other objects, superluminous SN I 2018ibb (a pair-instability SN candidate) and peculiar SN Ic 2022jli (the first example showing stable periodic modulation), show observational similarities to SNe Ic-CSM and may be categorized as SN Ic-CSM variants. Complemented with a large diversity in their light-curve evolution, we propose that SNe Ic-CSM (potentially linked to SNe Ibn/Icn) are a mixture of multiple channels that cover a range of properties in the progenitor star, the binary companion, and the binary orbit.
\end{abstract}

%\pagewiselinenumbers 

% for the page count (within 6 pages for the letter)
%\clearpage

\section{Introduction} \label{sec:intro}
Recent development in transient observations has revealed a rich diversity of a transient zoo (e.g., \cite{schulze2025}). `SN Ic-CSM' is a recently found class of objects, characterized by late-time emergence of narrow ($<$ a few 1,000 km s$^{-1}$) emission lines of elements formed in the oxygen core of a massive star (e.g., O and Mg) \citep{kuncarayakti2022}\footnote{SN Ib 2014C \citep{margutti2017} and SN Ic 2017dio \citep{kuncarayakti2018} showed transition to type IIn with strong Balmer emissions in the late-phase. In this paper, they are not classified as SN Ib-CSM or Ic-CSM, which we define are characterized by narrow emission lines in the H-poor environment.}. Within a still very limited sample, the best studied cases, SNe 2021ocs and 2022xxf, both showed a transition from type Ic (an explosion of a C+O star) to SN Ic-CSM \citep{kuncarayakti2023}, establishing their association with C+O star progenitors. In addition to the spectral evolution, they are distinct in the light-curve (LC) evolution from SNe Ic. The interaction between the SN ejecta and CSM (Circumstellar Interaction: CSI), both rich in oxygen, has been proposed as a power source.

The progenitor channel toward SNe Ic-CSM has not been clarified; it is not even clear whether SNe Ic-CSM form a single population. Also, a relation between SNe Ic-CSM and Ibn/Icn (intense interaction with H-deficient CSM only in the time scale of weeks; \cite{pastorello2007,gal-yam2022}) has not been clarified. 

SN 2022esa was discovered on 2022 March 12 (MJD 59650.61)\footnote{The date is shown in UT unless mentioned.} by Asteroid Terrestrial-impact Last Alert System (ATLAS) \citep{tonry2022}, in 2MFGC 13525 at $z=0.02314$. It was classified as SN Ia-CSM (H-rich CSI-powered SN Ia; \cite{dilday2012}) by \citet{lu2022}. In this Letter, we present follow-up observations of SN 2022esa. Its spectral evolution reveals that SN 2022esa is indeed a new member of the SN Ic-CSM class\footnote{Recently, \citet{griffith2025} also mentioned that it is a peculiar SN Ic.}. Its LC also shows unique features, including periodic modulation. We present these characteristic observational features, and discuss its main power source and the progenitor system.

\section{Observations and Data Reduction}\label{sec:obs}
Spectra of SN 2022esa were obtained on 2022 May 28 ($t=75$ d\footnote{In this Letter, we define the phase ($t$) of SN 2022esa as measured from the discovery date. The timescale (e.g., LCs shown in figures) is shown in the SN rest frame, corrected for redshift, for all the objects.}) and June 2 ($t=80$ d), using Kyoto Okayama Optical Low-dispersion Spectrograph with optical-fiber Integral Field Unit (KOOLS-IFU; \cite{matsubayashi2019}) mounted on the Kyoto University 3.8m Seimei Telescope \citep{kurita2020}. We used the VPH-blue grism with the wavelength coverage of $\sim 4000-8500$\AA\ and the spectral resolution of $\sim 600$. The total exposure time was 2,700 s in each night. A late-time spectrum was obtained on 2023 June 13 ($t=448$ d), using Faint Object Camera And Spectrograph (FOCAS) mounted on the Subaru telescope. We used a 0.8" slit and the B300 grism with no order-cut filter, covering 3650–8300\AA\ with the spectral resolution of $\sim 500$. The total exposure time was 3,600 s. Additionally, we obtained $V$ (120 s) and $R$-band (180 s) images on the same night with FOCAS.

The reduction of the FOCAS data was performed in the standard manner with IRAF\footnote{IRAF is distributed by the National Optical Astronomy Observatory, which is operated by the Association of Universities for Research in Astronomy (AURA) under a cooperative agreement with the National Science Foundation.} (see \cite{maeda2022_Iax}). The reduction of the KOOLS-IFU data follows similar procedures, with the Hydra package \citep{hydra} and custom routines\footnote{http://www.o.kwasan.kyoto-u.ac.jp/inst/p-kools/reduction-201806/index.html}.

\begin{figure}
\begin{center}
\includegraphics[width=1.0\columnwidth]{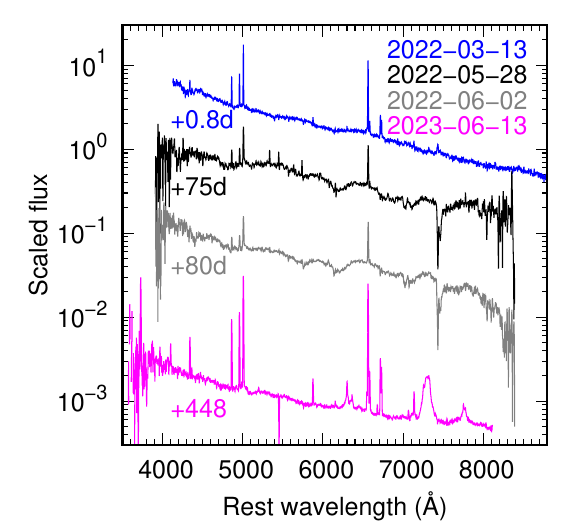}
\end{center}
\caption{The spectral evolution of SN 2022esa, including the one (0.8 d) reported by \citep{lu2022}. For the FOCAS spectrum on 448 d, the original (`SN+HII') spectrum is shown. 
%{Alt text: The figure shows spectra of SN 2022esa in time sequence, on 0.8, 75, 80, 448 days since the discovery from top to bottom. For day 448, the `SN+HII' spectrum is shown in magenta, while the `SN' spectrum is shown in red. The `SN' spectrum clearly shows emission lines from IMEs, i.e., the SN Ic-CSM features.}
}
\label{fig:spec}
\end{figure}

The spectra of SN 2022esa (figure \ref{fig:spec}) show a number of narrow emission lines from an underlying unresolved HII region. For the FOCAS spectrum, the contamination is especially substantial given the decreasing SN flux, and thus we subtracted it with a spectrum of a nearby HII region taken in the same 2D image. The flux was converted to the absolute scale by assuming the distance modulus of 35.24 mag and the Milky-Way (MW) extinction with $E(B-V) = 0.6$ mag. We assume that the extinction within the host is negligible; in addition to the absence of strong Na ID absorption, the Balmer decrement measured from the underlying HII region is consistent with the MW-only extinction. 

\begin{figure*}
\begin{center}
\includegraphics[width=\columnwidth]{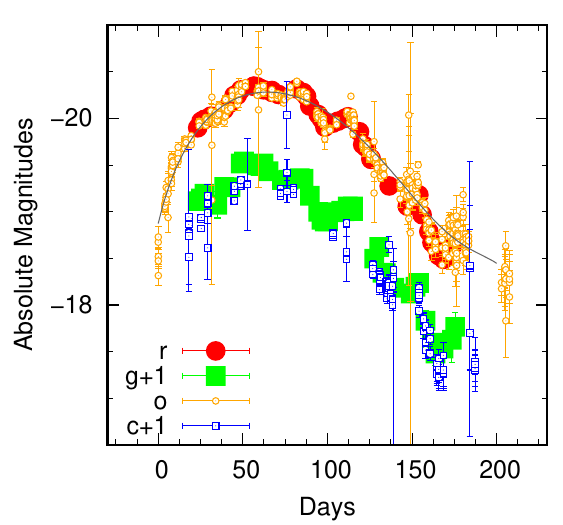}
\includegraphics[width=\columnwidth]{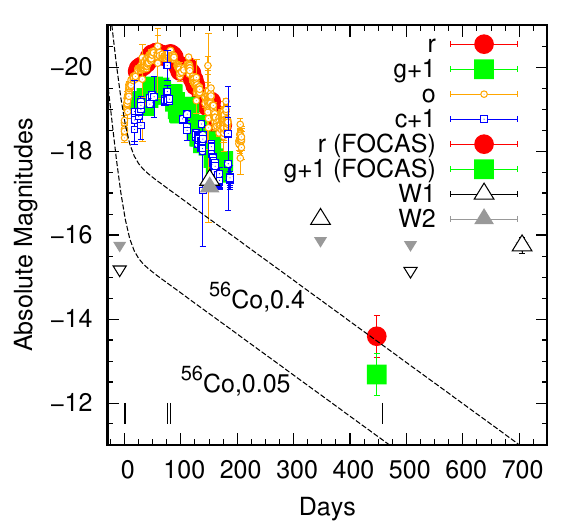}
\end{center}
\caption{The multi-band LCs of SN 2022esa around the peak (left) and those including the late-time measurements (right). The $g$- and $c$-band magnitudes are shifted by $+1$ mag for presentation. On the left panel, the mean LC (for ATLAS-o) is shown by a gray line. On the right panel, the WISE W1 (3.4$\micron$) and W2 (4.6 $\micron$) magnitudes (triangles) and $3\sigma$ upper limits (inversed triangles) are shown. Also shown are the full-deposition $^{56}$Ni/Co/Fe decay curve for $M$($^{56}$Ni)$=0.4$ and $0.05 M_\odot$, assuming that the explosion date is 15 days before the discovery. 
The epochs when the spectra were taken are indicated by short vertical lines on the $x$-axis.
%{Alt text: The left panel shows the light curves of SN 2022esa up to 250 days in the $r$, $g$, $o$, and $c$ bands. The right panel is the same but extended to 800 days, including the late-time $r$ and $g$ band magnitudes on day 448 as well as the WISE W1 and W2 magnitudes. }
}
\label{fig:lc}
\end{figure*}

The light curves (LCs) of SN 2022esa (figure \ref{fig:lc}) were constructed as follows. For the early phase up to $t \sim 200$ days, we used public resources. The ATLAS $o$ and $c$-band magnitudes were obtained using ATLAS forced photometry server\footnote{https://fallingstar-data.com/forcedphot/} \citep{tonry2018,shingles2021}. The ZTF $g$ and $r$-band magnitudes were obtained through the ALeRCE explorer\footnote{https://alerce.science/} \citep{foster2021}. We used the FOCAS spectrum to estimate the `SN-only' magnitude in the late phase ($t = 448$ d). First, spectrophotometry on the `SN + HII region' resulted in $V \sim 20.8$ and $R \sim 20.3$ mag (in Vega), being consistent with the photometry performed on the $V$ and $R$ images ($V = 21.10 \pm 0.12$ and $R = 20.29 \pm 0.14$ mag). We then performed spectrophotometry on the HII region-subtracted spectrum, obtaining the pure-SN magnitudes of $g = 23.6 \pm 0.3$ and $r = 23.3 \pm 0.3$ mag (in AB). 

We also constructed infrared LCs using the data from the NEOWISE mission \citep{mainzer2014} in the W1 (3.4 µm) and W2 (4.6 µm) bands, taken by the Wide-field Infrared Survey Explorer (WISE; \cite{wright2010}). We performed point-spread function (PSF) photometry at the position of the source in the NEOWISE difference images \citep{de2020,de2023}.

\section{Results}\label{sec:result}
\subsection{Modulation in the Light Curve}\label{sec:modulation}
\begin{figure*}
\begin{center}
\includegraphics[width=0.65\columnwidth]{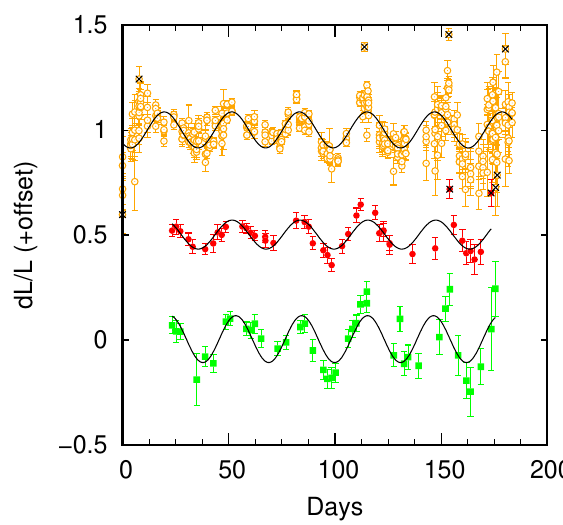}
\includegraphics[width=0.65\columnwidth]{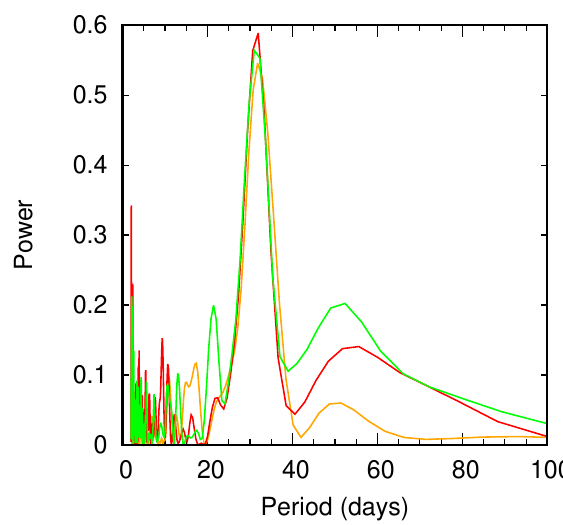}
\includegraphics[width=0.65\columnwidth]{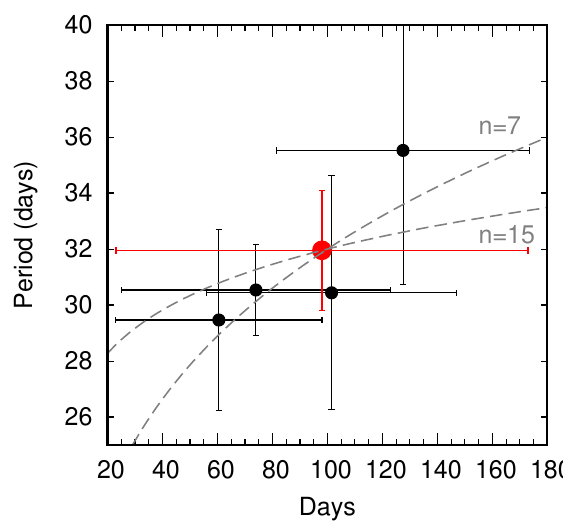}
\end{center}
\caption{
The fractional-residual LCs in the ATLAS-$o$ (orange), ZTF-$r$ (red), ZTF-$g$ (green), and the best fit sine curves through the LS analyses, are shown in the left panel. The data points that were rejected through the LS analyses are shown by crosses. In the middle panel, the power spectra, using the same color coordinates with those used in the left panel, are shown. In the right panel, the periods obtained through the sliding-window analysis (black) as well as using the full time window (red) are shown for the ZTF-$r$ data. The CSI-model predictions for the period change are shown by the dashed lines (see the main text). 
%{Alt text: The left panel shows three fractional-residual LCs; Atlas-$o$, ZTF-$r$, and TZF-$g$ from top to bottom. The middle panel shows the power spectra (the period on the $x$-axis and the power on the $y$-axis) in the three bands, showing the peak at $\sim 32$ days in all the bands. The right panel shows the derived period in each time window through the sliding--window analysis (4 points for the sliding analysis and 1 point for the full time window), shows as points with errors in the $x$-axis (the width of the window) and $y$-axis (the errors in the derived periods).}
}
\label{fig:period}
\end{figure*}

Figure \ref{fig:lc} shows bumpy structures in the LC evolution, the characteristic that is rarely seen in SNe. 
The evolution is coherent in different bands; 
excluding the Atlas $c$-band data that are not well sampled, we performed the periodgram analysis. We first fit the overall LC of a given band by a polynomial function of 5th order, 
creating the mean LC. This mean LC was subtracted from the original LC, and then the residual was further divided by the mean LC. The resulting `fractional-residual' LCs for the atlas $o$-band and ZTF $r$- and $g$-bands are shown in figure \ref{fig:period} (left). Surprisingly, SN 2022esa showed clear periodicity over multiple cycles, which has been reported previously only for peculiar SN Ic 2022jli \citep{moore2023,chen2024}\footnote{Another example, type I superluminous SN (SLSN-I) 2024afav, has recently been reported to show a quasi-periodic LC \citep{farah2025}. See also \citet{west2023} for SLSN-I 2020qlb.}. 

We performed the Lomb-Scargle (LS) periodogram analyses for the residual LCs separately in the $o$-, $r$-, and $g$-band data, using LombScargle module implemented in astropy. This was done iteratively by rejecting outliers by 3$\sigma$ clipping in the data-model residual. A sharp peak in the power spectrum is found at the period that is consistent among all the three bands; $31.8 \pm 2.8$ days ($o$), $32.0 \pm 2.1$ days ($r$), and $31.0 \pm 2.2$ days ($g$) (figure \ref{fig:period}, middle). To further investigate possible quasi-periodic behavior, we performed the sliding-window analysis with a window size of 100 days and a step size of 25 days on the ZTF-$r$ data. As shown in figure \ref{fig:period} (right), the data are consistent with no change in the period within 1$\sigma$, while there is a hint of a possible increase in the period over time\footnote{The sliding-window analysis of the Atlas-$o$ data resulted in the same conclusion, with a larger error in the derived period.}.

In case of SN 2022jli, the origin of the modulation has not been conclusively clarified \citep{moore2023,cartier2024}, but the `post-SN' binary interaction between the compact object left behind the explosion and the bound companion star has been proposed \citep{chen2024}. As compared to SN 2022jli, the modulation seen here has several distinct properties. SN 2022jli showed (1) a shorter period (12.4 d), and (2) an asymmetric temporal profile with a fast rise ($\leq 3$ d) while it is consistent with a symmetric profile for SN 2022esa. Further, (3) the modulation was observed in the late phase after $\sim 50$ d in SN 2022jli, while the early-rising LC of SN 2022esa is consistent with having the modulation already. There are also other differences between SNe 2022esa and 2022jli in the overall LC and spectral evolutions, which will be discussed in subsequent sections. 

Temporal variations are sometimes seen in radio LCs of stripped-envelope SNe (SESNe, i.e., SNe IIb/Ib/Ic) \citep{wellon2012}, and the quasi-periodic modulation has been reported for a few SNe \citep{ryder2004,kotak2006,soderberg2006}. The behavior has been generally interpreted to be caused by the modulation in the CSM density structure. Especially popular is the scenario that involves a luminous blue variable-like progenitor \citep{kotak2006}. Indeed, the temporal variation in the LC, while usually without periodic behavior, is common among SNe Ic-CSM (see section \ref{sec:lc}); the H-poor CSI in a non-smooth CSM distribution is a popular scenario \citep{kuncarayakti2023}.

\subsection{SN 2022esa as a new member of SNe Ic-CSM}\label{sec:spec}

SN 2022esa showed broad spectral features already in the early phase with no clear sign of H$\alpha$ from the SN (figure \ref{fig:spec}). The broad features were further developed at the maximum phase. Finally, in the late epoch, strong and narrow (but resolved) lines (FWHM $\sim 2000$ km s$^{-1}$) from intermediate mass elements (IMEs) such as O and Ca were clearly detected; these properties are inconsistent with SN Ia-CSM, but rather indicate that SN 2022esa was associated with O-rich environments, i.e., SN Ic or its variants. 

\begin{figure}
\begin{center}
\includegraphics[width=0.9\columnwidth]{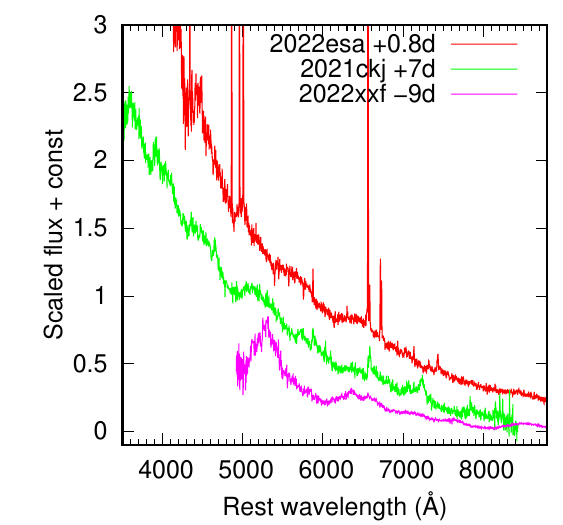}
\includegraphics[width=0.9\columnwidth]{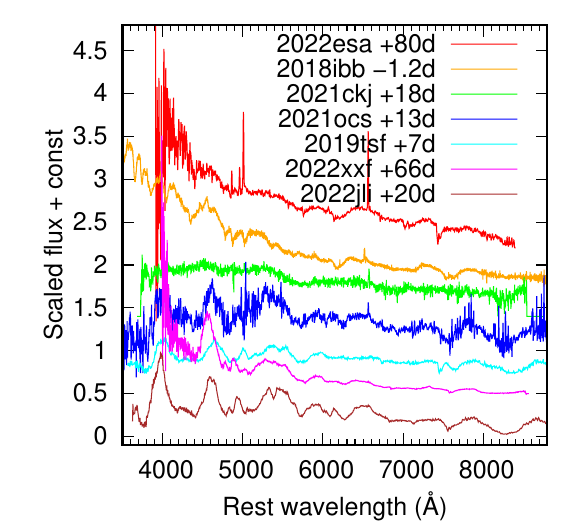}
\includegraphics[width=0.87\columnwidth]{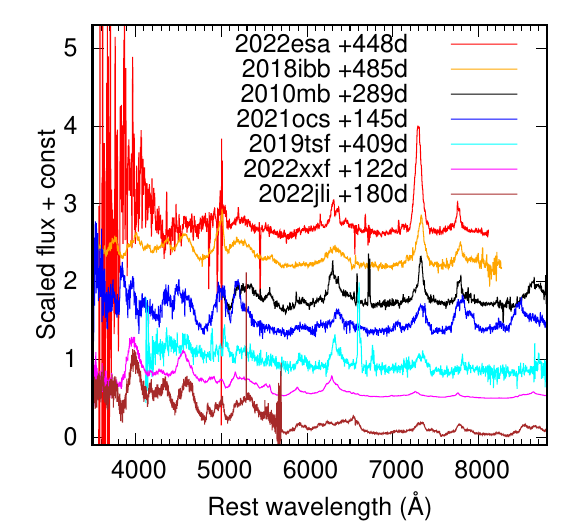}
\end{center}
\caption{Comparison of the spectra of SN 2022esa to SNe Ic-CSM and related objects, in the early (top), peak/intermediate (middle), and late phases (bottom). In the bottom panel, the HII region-subtracted spectrum is shown. The data of the comparison objects are from the literature in section \ref{sec:spec}, and downloaded through WISeREP \citep{wiserep2012} or provided by the authors of the literature. The phases for the comparison objects are measured from the peak light (in their rest frames).
%{Alt text: Comparison of the spectra of SN 2022esa with various supernovae, shown in three panels at different epochs. Three, seven, and seven spectra including SN 2022esa are shown in the top, middle, and bottom panels. }
}
\label{fig:spec_comp_Ic-CSM}
\end{figure}

We compare the spectral evolution of SN 2022esa with those of SNe Ic-CSM and a few related objects (figure \ref{fig:spec_comp_Ic-CSM}). 
SN Icn 2021ckj is a rapidly decaying SN where the interaction with `confined' C-rich CSM is the main energy source \citep{pellegrino2022,nagao2023}. SNe 2010mb \citep{ben-ami2014}, 2021ocs \citep{kuncarayakti2022}, and 2022xxf \citep{kuncarayakti2023} were classified as SNe Ic-CSM based on their late-time spectra.  
They showed flat or slowly decaying LCs even with multiple peaks. From these features, SNe Ic-CSM have been proposed to be SNe Ic exploded within dense and extended O-rich CSM. SNe 2021ocs and 2022xxf did not show the signatures of strong CSI in the early phase, and were classified as a canonical SN Ic and a broad-lined SN Ic (SN Ic-BL) initially, pointing to the diverse nature of the SN ejecta in the SN Ic-CSM class. SN 2019tsf is a peculiar SN Ib in its LC evolution \citep{sollerman2020}, later showing the spectral features of SNe Ic-CSM \citep{pyykkinen2025}. SLSN-I 2018ibb has been suggested to be a pair-instability SN (PISN) \citep{schulze2024,nagele2024}, with the late-time spectrum indicating an additional power provided by the H-poor CSI \citep{chugai2024}. Finally, SN Ib 2022jli, the first SN showing the clear periodicity, is also shown for comparison \citep{moore2023,chen2024}. 

SN 2022esa shows striking similarity to SNe Ic-CSM. The earliest spectrum is overall similar to the spectrum of SN 2022xxf at the first peak. 
The broad features also show resemblance to SN Icn 2021ckj, while the narrow emission lines from highly-ionized carbon are missing in SN 2022esa. These similarities indicate that SN 2022esa is related to an oxygen-star progenitor, i.e., `SN Ic'. 

The spectrum around the peak of SN 2022esa shows clear similarity to those of SNe Ic-CSM 2021ocs and 2022xxf. Interestingly, the candidate PISN 2018ibb and the `periodic' SN Ic 2022jli are similar to SN Ic-CSM objects and SN 2022esa, inferring a link of these peculiar objects to the SN Ic-CSM class; we suggest that they could indeed be variants of SNe Ic-CSM, potentially sharing a similar power source and/or spectral-formation process.

Finally, the late-time spectrum (+448 d) shows striking similarity to those of SNe Ic-CSM and (somehow surprisingly) SN 2018ibb, characterized by narrow emission lines from O and IMEs, together with the blue Fe complex. SN Ib 2019tsf starts showing the 'SN Ic-CSM' characteristics, and it might be classified as `SN Ib-CSM'. Interestingly, SN 2022jli also shows an overall similar spectrum to SNe Ic-CSM; it however shows important differences, i.e., broad H$\alpha$ emission instead of [O \emissiontype{I}]. 

\subsection{Possible power sources}\label{sec:lc}
The large luminosity ($< -20$ mag) and long duration ($< -19$ mag for $\sim 150$ days) indicate the radiated energy of SN 2022esa reaching at least $\sim 3 \times 10^{50}$ erg. In addition, the color remained rather constant and blue ($g-r$ and $c-o \sim 0$ mag) throughout the period up to $t \sim 500$ d. The CSI is a promising candidate to explain these properties. 
Figure \ref{fig:lc_comp} compares the LCs of SN 2022esa and related objects, including `H-rich CSI' SN IIn 2010jl \citep{zhang2012} and Ia-CSM 2020uem \citep{uno2023}. In addition to the lack of the H features in the spectra, the LC evolution also indicates that SN 2020esa is neither of the SN Ia-CSM nor SN IIn class\footnote{We however note that the LCs of SNe IIn are very diverse.}. 

Some SNe Ic-CSM show two or more peaks, which has been associated with two different energy sources -- $^{56}$Ni/Co/Fe heating at the first peak, and the O-rich CSI thereafter. The most likely energy source for SN 2022esa is thus the O-rich CSI, and the LC periodicity reflects the fluctuation in the CSM density. Under this scenario, we have two additional constraints; (1) the large peak luminosity indicates a very high mass-loss rate, reaching $\geq 0.01 - 0.1 M_\odot$ yr$^{-1}$ \citep{maeda2013_10jl,uno2023} just before the explosion (unlike SNe 2021ocs and 2022xxf showing the late-time emergence of the CSI), and (2) it requires a stable periodic phenomenon to introduce the CSM density fluctuation (unlike SN 2022xxf showing a bumpy LC but without clear periodicity). 

We note two additional hints supporting the CSI scenario: (1) The properties of the radio emission \citep{griffith2025}, $\sim 10^{28}$ erg s$^{-1}$ Hz$^{-1}$ at $\sim 1.5$ yrs at 6 GHz, is typical of SNe IIn \citep{chevalier2006,maeda2013_radio}. (2) IR excess is seen in the late phase (figure \ref{fig:lc}), similar to those frequently observed in SNe IIn. The IR emission signals dusty environments associated with dense CSM either through an echo from pre-existing CS dust (at a larger distance; \cite{maeda2015_echo}) or through the reprocessing of the optical photons to the IR by newly formed dust (at a smaller scale; \cite{maeda2013_10jl}). Given the blue optical color even in the late phase, the pre-existing dust is a more likely scenario. We will further investigate the nature of the IR excess in a separate paper.

If the pre-SN mass loss would be perfectly periodic with no change in the mass-loss velocity ($V_{\rm w}$), the CSM `shells' are equally spaced. Then, the observed LC modulation period is inversely proportional to the SN shock velocity $V_{\rm SN}$. Therefore, we expect that the period is constant (for constant $V_{\rm SN}$) or increasing (for decreasing $V_{\rm SN}$). For example, for the decelerated SN shock in a steady-state wind-like CSM, $V_{\rm SN} \propto t^{-1/(n-2)}$ where $n$ is the density slope in the SN ejecta ($\rho_{\rm SN} \propto v^{-n}$). Figure \ref{fig:period} (right) shows the predictions for $n= 7$ and $15$ covering a range of SN-ejecta properties \citep{chevalier1982}. 
The observed period is consistent with this picture (i.e., no change within the error, but with a hint of an increasing period).

The `post'-SN binary interaction scenario proposed for SN 2022jli, where the `central engine' through the post-SN binary interaction powers the LC periodically, is not rejected. However, the main argument against the CSI scenario for SN 2022jli, namely the variability shorter than the light-travel time, does not apply for SN 2022esa. Also, given its high luminosity, SN 2022esa would require a very strong binary interaction and a high mass-transfer rate, and the period may likely decrease rapidly, inconsistent with the constant or slowly increasing period in SN 2022esa.

\subsection{SN-site environment}\label{sec:environment}
We measured metallicity at the SN site and the nearby H II region, using the O3N2 and N2 calibrations from \citet{marino13}. We found that the metallicities for the two regions are the same within an error. Two calibrations provide consistent results; $12 + \rm{log(O/H)} \sim 8.3 \pm 0.2$ dex (N2) or $12 + \rm{log(O/H)} \sim 8.2 \pm 0.2$ dex (O3N2), i.e., $Z \sim 0.4 - 0.5 Z_\odot$. This is similar to the cases of some strongly interacting SNe, e.g. Type IIn SN 2010jl \citep{stoll11,maeda2013_10jl}.

The equivalent width of the H$\alpha$ emission line (H$\alpha$EW) is frequently used to infer the age of the stellar population and the lifetime of the progenitor star (see e.g. \cite{hk18-ifu}). The measured H$\alpha$EW of SN~2022esa environment is around 600\AA. Using the Starburst99 single-burst model \citep{leitherer99} at half solar metallicity, this corresponds to a stellar population age of around 4.7 Myr, thus to the lifetime of a $\sim50$ M$_\odot$ star. Such a massive star is expected to evolve to a massive WR star \citep{heger2003}. While the single-burst model can overestimate the progenitor mass, the young stellar population age suggests a massive star origin for SN 2022esa.

\section{Discussion}\label{sec:discussion}
\subsection{Progenitor System: WR-WR or WR-BH binary?}\label{sec:progenitor}
Under the CSI scenario, the LC covering $\sim 1$ yr reflects the activity of the progenitor (system) in the final $\sim (V_{\rm SN}/V_{\rm w})$ yrs $\sim 10$ yrs, 
where $V_{\rm SN} \sim 10,000$ km s$^{-1}$ and $V_{\rm w} \sim 1,000$ km s$^{-1}$ are assumed. 
The LC modulation period is translated into $\sim 1$ yr for the variability period of the pre-SN mass loss. A binary interaction, where one component is the C+O-star SN progenitor, is an appealing possibility to create such a stable periodicity.

The circularized close binary, which is believed to be a major channel toward SESNe \citep{fang2019}, would not create a stable periodicity. Alternatively, a highly eccentric binary experiences repeated binary interactions every time the system passes the peri center, potentially resulting in a periodic mass loss. This suggests that the progenitor of SN 2022esa likely evolved to a C+O star without strong binary interaction (which circularizes the orbit) -- a massive WR star that has lost the outer H and He envelopes by its strong wind. Assuming the total mass of the pre-SN binary systems as $\sim 20-30 M_\odot$, the average pre-SN binary separation must have been $\sim 500 R_\odot$. This again points to a massive WR star progenitor that can avoid becoming a red supergiant, thus not experiencing a Roche-lobe overflow and/or common envelope (CE) that will circularize the orbit and lead to a compact binary. 

No trace of hydrogen may suggest an H-deficient companion star, i.e., either a WR-WR system (analogous to eccentric WR-WR binary Apep; \cite{callingham2020}) or a WR-BH system, noting that a neutron-star (NS) companion is unlikely if the progenitor is a massive WR. While the (still rare) observed WR-BH systems generally show close circularized orbits \citep{prestwich2007,binder2021}, it may simply be due to observational biases; for O star-WR binaries (as a progenitor to WR-BH binaries) such as WR140, a period of a year to decades with an eccentric orbit is common (e.g., \cite{williams2009}). The ejecta mass of SN 2022esa would not exceed half of the total mass for the binary system considered here, and the binary likely survived after SN 2022esa. Therefore, the proposed WR-WR or WR-BH system will ultimately lead to the formation of a BH-BH binary. 

\begin{figure}
\begin{center}
\includegraphics[width=\columnwidth]{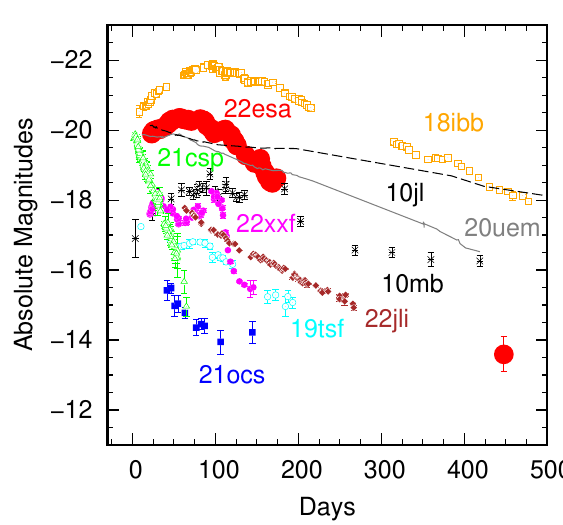}
\end{center}
\caption{The r-band LC of SN 2022esa as compared to SNe Icn and Ic-CSM (and related objects). The color coordinates are the same with figure \ref{fig:spec_comp_Ic-CSM} (note that SN Icn 2021ckj is replaced by SN Icn 2021csp; \cite{perley2022}). Also shown here are SNe Ia-CSM 2020uem and IIn 2010jl. 
%{Alt text: The r-band light curve of SN 2022esa covering 500 days, together with those of other nine supernovae. }
}
\label{fig:lc_comp}
\end{figure}

The LC decay of SN 2022esa is much faster than SNe interacting with extended CSM like SNe 2010jl and 2020uem. This indicates a steep CSM density gradient and an increasing mass-loss rate toward the SN. 
The binary interaction was thus `switched on' in the last decade before the SN. Such a synchronization requires that the final evolution of the progenitor drove the binary interaction. One possibility is inflation of the progenitor star or some sort of activity in the final phase \citep{ouchi2019} that could lead to an increasingly strong binary interaction toward the SN. For example, a massive WR progenitor has been proposed for SNe Ibn/Icn \citep{pastorello2007,gal-yam2022} -- putting such a star in an eccentric binary would lead to further enhancement of the mass loss and could lead to objects like SN 2022esa. 

In the `post-SN' binary interaction scenario, the observed period reflects the post-SN binary period -- the average separation of $\sim 100 R_\odot$.  
Given the lack of H$\alpha$, the companion star would also be a C+O star; it is unlikely that both stars evolve to C+O stars through strong binary interaction, and a massive WR-WR binary is a favored system also in this scenario. A few caveats should be noted; the emergence of H$\alpha$ might be a transient feature and could simply be missed, or the H$\alpha$ may be hidden as the binary interaction happens deep inside the ejecta. In these cases, a WR-O/B star binary progenitor system would also be possible. Either way, again the system may ultimately evolve to a BH-BH binary. 

\subsection{SN Ic-CSM diversities: Multiple populations?}\label{sec:diversity}
The speed of the spectral evolution is diverse among SNe Ic-CSM. Also, the early-phase spectra show some diversities, e.g., SNe 2021ocs and 2022xxf classified as normal Ic and Ic-BL. While overall features are shared in the late phase, there are also differences; in figure \ref{fig:spec_comp_Ic-CSM}, only SN 2022jli shows H$\alpha$ and a lack of [O \emissiontype{I}]6300/6363\footnote{It is thus not SN Ic-CSM. We however note the emergence of the [O \emissiontype{I}] in the later epoch \citep{cartier2024} as another indication of its link to SNe Ic-CSM.}. It is also seen that SN 2018ibb shows broader features than the others. 

SNe Ic-CSM shows a huge diversity in the LC evolution (figure \ref{fig:lc_comp}); the peak magnitude, the time scale, and the evolution morphology are diverse. This may indicate that SNe Ic-CSM are likely a mixture of different populations originated in multiple progenitor channels. For most of them, the O-rich CSI is likely the main power source, with diverse properties in the CSM (and thus in the mass-loss history). 
Even the underlying power source may not be unique, possibly including the post-SN binary interaction. These channels may include a massive WR-WR or WR-BH binary (proposed here for SN 2022esa), a standard (relatively low-mass) SESN binary channel but extreme conditions for the binary mass transfer (perhaps involving a CE), a WR + normal (perhaps O) star binary (proposed for SN 2022jli), and a PISN (proposed for 2018ibb). 
In most cases, the binarity seems to be a key; depending on the nature of the binary components (types of stars and masses) and the orbit (separation and eccentricity), a huge diversity within the `SN Ic-CSM' class (and possibly including Ibn/Icn) may be created.

\section{Concluding Remarks}\label{sec:concluding}
In this Letter, we presented unique features of SN 2022esa. In addition to the bright optical emission, it showed a stable periodicity ($\sim 32$ d) in the LC, with a possible hint of a slowly increasing period over $\sim 200$ days. Further, the spectra evolved into the characteristic `SN Ic-CSM' late-time spectrum, showing blue Fe bumps and narrow emission lines from IMEs. 

For the power source, both the CSI and the post-SN binary interactions remain possible, while the CSI is more likely. In either case, we suggest a massive WR progenitor star, in a binary with either another WR or a BH. For the CSI scenario, we further require a highly eccentric orbit. Therefore, SN 2022esa may represent a progenitor system that will eventually evolve to a BH-BH binary. 

We showed that the SNe Ic-CSM class has a huge diversity, pointing out that SNe 2018ibb (a PISN candidate) and 2022jli (another example showing a clear periodicity) have similarities to SNe Ic-CSM. We suggest that SNe Ic-CSM represent a mixture of multiple populations. The progenitor mass may span from the `canonical SESN progenitor' ($M_{\rm ZAMS} < 20 M_\odot$) to a massive WR star ($> 20 M_\odot$), or even to the PISN regime ($> 100 M_\odot$). Furthermore, for many of them, a binary interaction seems to play a key role, adding a source of the diversity depending on the binary configuration that may further connect them to SNe Ibn/Icn. 

SNe Ic-CSM can thus represent a rich and diverse zoo of massive binary evolution, including progenitors toward BH-BH binaries. As such, further studying them, both in observationally and theoretically, will provide a major contribution to uncover the still-unclarified final evolution of massive stars, roles of binarity, and formation channels of double compact-object binaries.

\begin{ack}
The authors thank Kentaro Aoki for his excellent assistance for the observation with the Subaru telescope (under the program S23A-023), and Niko Pyykkinen for providing the unpublished spectrum of SN 2019tsf. The data from the Seimei telescope were obtained under the KASTOR (Kanata And Seimei Transient Observation Regime) project (under the programs 22A-N-CT09 and 22A-K-0030). The Seimei telescope is jointly operated by Kyoto University and the Astronomical Observatory of Japan (NAOJ), with assistance provided by the Optical and Near-Infrared Astronomy Inter-University Cooperation Program.
Some public data are obtained from WISeREP (https://www.wiserep.org).

\end{ack}

\section*{Funding}
K.M. acknowledges support from JSPS KAKENHI grant (JP24KK0070, JP24H01810, JP24K00682, and JP20H00174). 
H.K. was funded by the Research Council of Finland projects 324504, 328898, and 353019. The work is partly supported by the JSPS Open Partnership Bilateral Joint Research Projects between Japan and Finland (K.M and H.K; JPJSBP120229923). 

\section*{Data availability} 
 The data underlying this article will be available through WiseRep.  
 
% Sample Data Availability Statements 
% https://academic.oup.com/pages/open-research/research-data#Data%20Availability%20Statements

%\appendix %%%%%%%%%%%%%%%%%%%%%%%%%%%%%%%%%%%%%%%%%%%%%%%%%%%%%%%%
%\section*{Case of single paragraph}
% No section number is necessary. Add ``*'' after \verb/\section/.
%
%%%%% 
%\section{Case of two or more paragraphs}
%
% Text of appendix
%
%\section{Case of two or more paragraphs}
%
% Text of appendix

% Any journal's BST file (e.g., apj.bst) can be used as PASJ's BST is unavailable.    
% \bibliographystyle{****}
% \bibliography{****}
\bibliography{sn2022esa_pasj}{}

@ARTICLE{wiserep2012,
       author = {{Yaron}, Ofer and {Gal-Yam}, Avishay},
        title = "{WISeREP{\textemdash}An Interactive Supernova Data Repository}",
      journal = {\pasp},
         year = 2012,
        month = jul,
       volume = {124},
       number = {917},
        pages = {668},
          doi = {10.1086/666656},
archivePrefix = {arXiv},
       eprint = {1204.1891},
 primaryClass = {astro-ph.IM},
       adsurl = {https://ui.adsabs.harvard.edu/abs/2012PASP..124..668Y}
}

@ARTICLE{west2023,
       author = {{West}, S.~L. and {Lunnan}, R. and {Omand}, C.~M.~B. and {Kangas}, T. and {Schulze}, S. and {Strotjohann}, N.~L. and {Yang}, S. and {Fransson}, C. and {Sollerman}, J. and {Perley}, D. and {Yan}, L. and {Chen}, T.-W. and {Chen}, Z.~H. and {Taggart}, K. and {Fremling}, C. and {Bloom}, J.~S. and {Drake}, A. and {Graham}, M.~J. and {Kasliwal}, M.~M. and {Laher}, R. and {Medford}, M.~S. and {Neill}, J.~D. and {Riddle}, R. and {Shupe}, D.},
        title = "{SN 2020qlb: A hydrogen-poor superluminous supernova with well-characterized light curve undulations}",
      journal = {\aap},
         year = 2023,
        month = feb,
       volume = {670},
          eid = {A7},
        pages = {A7},
          doi = {10.1051/0004-6361/202244086},
archivePrefix = {arXiv},
       eprint = {2205.11143},
 primaryClass = {astro-ph.HE},
       adsurl = {https://ui.adsabs.harvard.edu/abs/2023A&A...670A...7W}
}

@ARTICLE{ben-ami2014,
       author = {{Ben-Ami}, Sagi and {Gal-Yam}, Avishay and {Mazzali}, Paolo A. and {Gnat}, Orly and {Modjaz}, Maryam and {Rabinak}, Itay and {Sullivan}, Mark and {Bildsten}, Lars and {Poznanski}, Dovi and {Yaron}, Ofer and {Arcavi}, Iair and {Bloom}, Joshua S. and {Horesh}, Assaf and {Kasliwal}, Mansi M. and {Kulkarni}, Shrinivas R. and {Nugent}, Peter E. and {Ofek}, Eran O. and {Perley}, Daniel and {Quimby}, Robert and {Xu}, Dong},
        title = "{SN 2010mb: Direct Evidence for a Supernova Interacting with a Large Amount of Hydrogen-free Circumstellar Material}",
      journal = {\apj},
         year = 2014,
        month = apr,
       volume = {785},
       number = {1},
          eid = {37},
        pages = {37},
          doi = {10.1088/0004-637X/785/1/37},
archivePrefix = {arXiv},
       eprint = {1309.6496},
 primaryClass = {astro-ph.SR},
       adsurl = {https://ui.adsabs.harvard.edu/abs/2014ApJ...785...37B}
}

@ARTICLE{binder2021,
       author = {{Binder}, Breanna A. and {Sy}, Janelle M. and {Eracleous}, Michael and {Christodoulou}, Dimitris M. and {Bhattacharya}, Sayantan and {Cappallo}, Rigel and {Laycock}, Silas and {Plucinsky}, Paul P. and {Williams}, Benjamin F.},
        title = "{The Wolf-Rayet + Black Hole Binary NGC 300 X-1: What is the Mass of the Black Hole?}",
      journal = {\apj},
         year = 2021,
        month = mar,
       volume = {910},
       number = {1},
          eid = {74},
        pages = {74},
          doi = {10.3847/1538-4357/abe6a9},
archivePrefix = {arXiv},
       eprint = {2102.07065},
 primaryClass = {astro-ph.HE},
       adsurl = {https://ui.adsabs.harvard.edu/abs/2021ApJ...910...74B}
}

@ARTICLE{callingham2020,
       author = {{Callingham}, J.~R. and {Crowther}, P.~A. and {Williams}, P.~M. and {Tuthill}, P.~G. and {Han}, Y. and {Pope}, B.~J.~S. and {Marcote}, B.},
        title = "{Two Wolf-Rayet stars at the heart of colliding-wind binary Apep}",
      journal = {\mnras},
         year = 2020,
        month = jan,
       volume = {495},
       number = {3},
        pages = {3323-3331},
          doi = {10.1093/mnras/staa1244},
archivePrefix = {arXiv},
       eprint = {2005.00531},
 primaryClass = {astro-ph.SR},
       adsurl = {https://ui.adsabs.harvard.edu/abs/2020MNRAS.495.3323C}
}

@ARTICLE{chen2024,
       author = {{Chen}, Ping and {Gal-Yam}, Avishay and {Sollerman}, Jesper and {Schulze}, Steve and {Post}, Richard S. and {Liu}, Chang and {Ofek}, Eran O. and {Das}, Kaustav K. and {Fremling}, Christoffer and {Horesh}, Assaf and {Katz}, Boaz and {Kushnir}, Doron and {Kasliwal}, Mansi M. and {Kulkarni}, Shri R. and {Liu}, Dezi and {Liu}, Xiangkun and {Miller}, Adam A. and {Rose}, Kovi and {Waxman}, Eli and {Yang}, Sheng and {Yao}, Yuhan and {Zackay}, Barak and {Bellm}, Eric C. and {Dekany}, Richard and {Drake}, Andrew J. and {Fang}, Yuan and {Fynbo}, Johan P.~U. and {Groom}, Steven L. and {Helou}, George and {Irani}, Ido and {Jegou du Laz}, Theophile and {Liu}, Xiaowei and {Mazzali}, Paolo A. and {Neill}, James D. and {Qin}, Yu-Jing and {Riddle}, Reed L. and {Sharon}, Amir and {Strotjohann}, Nora L. and {Wold}, Avery and {Yan}, Lin},
        title = "{A 12.4-day periodicity in a close binary system after a supernova}",
      journal = {\nat},
         year = 2024,
        month = jan,
       volume = {625},
       number = {7994},
        pages = {253-258},
          doi = {10.1038/s41586-023-06787-x},
archivePrefix = {arXiv},
       eprint = {2310.07784},
 primaryClass = {astro-ph.HE},
       adsurl = {https://ui.adsabs.harvard.edu/abs/2024Natur.625..253C}
}

@ARTICLE{chevalier2006,
       author = {{Chevalier}, Roger A. and {Fransson}, Claes and {Nymark}, Tanja K.},
        title = "{Radio and X-Ray Emission as Probes of Type IIP Supernovae and Red Supergiant Mass Loss}",
      journal = {\apj},
         year = 2006,
        month = apr,
       volume = {641},
       number = {2},
        pages = {1029-1038},
          doi = {10.1086/500528},
archivePrefix = {arXiv},
       eprint = {astro-ph/0509468},
 primaryClass = {astro-ph},
       adsurl = {https://ui.adsabs.harvard.edu/abs/2006ApJ...641.1029C}
}

@ARTICLE{chugai2024,
       author = {{Chugai}, N.~N.},
        title = "{Superluminous Supernova SN 2018ibb: Circumstellar Shell and Spectral Effects}",
      journal = {Astronomy Letters},
         year = 2024,
        month = aug,
       volume = {50},
       number = {8},
        pages = {502-509},
          doi = {10.1134/S1063773724700361},
archivePrefix = {arXiv},
       eprint = {2410.17580},
 primaryClass = {astro-ph.HE},
       adsurl = {https://ui.adsabs.harvard.edu/abs/2024AstL...50..502C}
}

@ARTICLE{dilday2012,
       author = {{Dilday}, B. and {Howell}, D.~A. and {Cenko}, S.~B. and {Silverman}, J.~M. and {Nugent}, P.~E. and {Sullivan}, M. and {Ben-Ami}, S. and {Bildsten}, L. and {Bolte}, M. and {Endl}, M. and {Filippenko}, A.~V. and {Gnat}, O. and {Horesh}, A. and {Hsiao}, E. and {Kasliwal}, M.~M. and {Kirkman}, D. and {Maguire}, K. and {Marcy}, G.~W. and {Moore}, K. and {Pan}, Y. and {Parrent}, J.~T. and {Podsiadlowski}, P. and {Quimby}, R.~M. and {Sternberg}, A. and {Suzuki}, N. and {Tytler}, D.~R. and {Xu}, D. and {Bloom}, J.~S. and {Gal-Yam}, A. and {Hook}, I.~M. and {Kulkarni}, S.~R. and {Law}, N.~M. and {Ofek}, E.~O. and {Polishook}, D. and {Poznanski}, D.},
        title = "{PTF 11kx: A Type Ia Supernova with a Symbiotic Nova Progenitor}",
      journal = {Science},
         year = 2012,
        month = aug,
       volume = {337},
       number = {6097},
        pages = {942},
          doi = {10.1126/science.1219164},
archivePrefix = {arXiv},
       eprint = {1207.1306},
 primaryClass = {astro-ph.CO},
       adsurl = {https://ui.adsabs.harvard.edu/abs/2012Sci...337..942D}
}

@ARTICLE{fang2019,
       author = {{Fang}, Qiliang and {Maeda}, Keiichi and {Kuncarayakti}, Hanindyo and {Sun}, Fengwu and {Gal-Yam}, Avishay},
        title = "{A hybrid envelope-stripping mechanism for massive stars from supernova nebular spectroscopy}",
      journal = {Nature Astronomy},
         year = 2019,
        month = mar,
       volume = {3},
        pages = {434-439},
          doi = {10.1038/s41550-019-0710-6},
archivePrefix = {arXiv},
       eprint = {1808.04834},
 primaryClass = {astro-ph.HE},
       adsurl = {https://ui.adsabs.harvard.edu/abs/2019NatAs...3..434F}
}

@ARTICLE{foster2021,
       author = {{F{\"o}rster}, F. and {Cabrera-Vives}, G. and {Castillo-Navarrete}, E. and {Est{\'e}vez}, P.~A. and {S{\'a}nchez-S{\'a}ez}, P. and {Arredondo}, J. and {Bauer}, F.~E. and {Carrasco-Davis}, R. and {Catelan}, M. and {Elorrieta}, F. and {Eyheramendy}, S. and {Huijse}, P. and {Pignata}, G. and {Reyes}, E. and {Reyes}, I. and {Rodr{\'\i}guez-Mancini}, D. and {Ruz-Mieres}, D. and {Valenzuela}, C. and {{\'A}lvarez-Maldonado}, I. and {Astorga}, N. and {Borissova}, J. and {Clocchiatti}, A. and {De Cicco}, D. and {Donoso-Oliva}, C. and {Hern{\'a}ndez-Garc{\'\i}a}, L. and {Graham}, M.~J. and {Jord{\'a}n}, A. and {Kurtev}, R. and {Mahabal}, A. and {Maureira}, J.~C. and {Mu{\~n}oz-Arancibia}, A. and {Molina-Ferreiro}, R. and {Moya}, A. and {Palma}, W. and {P{\'e}rez-Carrasco}, M. and {Protopapas}, P. and {Romero}, M. and {Sabatini-Gacitua}, L. and {S{\'a}nchez}, A. and {San Mart{\'\i}n}, J. and {Sep{\'u}lveda-Cobo}, C. and {Vera}, E. and {Vergara}, J.~R.},
        title = "{The Automatic Learning for the Rapid Classification of Events (ALeRCE) Alert Broker}",
      journal = {\aj},
         year = 2021,
        month = may,
       volume = {161},
       number = {5},
          eid = {242},
        pages = {242},
          doi = {10.3847/1538-3881/abe9bc},
archivePrefix = {arXiv},
       eprint = {2008.03303},
 primaryClass = {astro-ph.IM},
       adsurl = {https://ui.adsabs.harvard.edu/abs/2021AJ....161..242F}
}

@ARTICLE{gal-yam2022,
       author = {{Gal-Yam}, A. and {Bruch}, R. and {Schulze}, S. and {Yang}, Y. and {Perley}, D.~A. and {Irani}, I. and {Sollerman}, J. and {Kool}, E.~C. and {Soumagnac}, M.~T. and {Yaron}, O. and {Strotjohann}, N.~L. and {Zimmerman}, E. and {Barbarino}, C. and {Kulkarni}, S.~R. and {Kasliwal}, M.~M. and {De}, K. and {Yao}, Y. and {Fremling}, C. and {Yan}, L. and {Ofek}, E.~O. and {Fransson}, C. and {Filippenko}, A.~V. and {Zheng}, W. and {Brink}, T.~G. and {Copperwheat}, C.~M. and {Foley}, R.~J. and {Brown}, J. and {Siebert}, M. and {Leloudas}, G. and {Cabrera-Lavers}, A.~L. and {Garcia-Alvarez}, D. and {Marante-Barreto}, A. and {Frederick}, S. and {Hung}, T. and {Wheeler}, J.~C. and {Vink{\'o}}, J. and {Thomas}, B.~P. and {Graham}, M.~J. and {Duev}, D.~A. and {Drake}, A.~J. and {Dekany}, R. and {Bellm}, E.~C. and {Rusholme}, B. and {Shupe}, D.~L. and {Andreoni}, I. and {Sharma}, Y. and {Riddle}, R. and {van Roestel}, J. and {Knezevic}, N.},
        title = "{A WC/WO star exploding within an expanding carbon-oxygen-neon nebula}",
      journal = {\nat},
         year = 2022,
        month = jan,
       volume = {601},
       number = {7892},
        pages = {201-204},
          doi = {10.1038/s41586-021-04155-1},
archivePrefix = {arXiv},
       eprint = {2111.12435},
 primaryClass = {astro-ph.SR},
       adsurl = {https://ui.adsabs.harvard.edu/abs/2022Natur.601..201G}
}

@ARTICLE{griffith2025,
       author = {{Griffith}, Olivia and {Showerman}, Grace and {Sarbadhicary}, Sumit K. and {Harris}, Chelsea E. and {Chomiuk}, Laura and {Sollerman}, Jesper and {Lundqvist}, Peter and {Mold{\'o}n}, Javier and {Torres}, Miguel P{\'e}rez- and {Kool}, Erik C. and {Moriya}, Takashi J.},
        title = "{A Late-time Radio Survey of Type Ia-CSM Supernovae with the Very Large Array}",
      journal = {\apj},
         year = 2025,
        month = dec,
       volume = {995},
       number = {1},
          eid = {54},
        pages = {54},
          doi = {10.3847/1538-4357/ae17b0},
archivePrefix = {arXiv},
       eprint = {2506.19071},
 primaryClass = {astro-ph.HE},
       adsurl = {https://ui.adsabs.harvard.edu/abs/2025ApJ...995...54G}
}

@ARTICLE{heger2003,
       author = {{Heger}, A. and {Fryer}, C.~L. and {Woosley}, S.~E. and {Langer}, N. and {Hartmann}, D.~H.},
        title = "{How Massive Single Stars End Their Life}",
      journal = {\apj},
         year = 2003,
        month = jul,
       volume = {591},
       number = {1},
        pages = {288-300},
          doi = {10.1086/375341},
archivePrefix = {arXiv},
       eprint = {astro-ph/0212469},
 primaryClass = {astro-ph},
       adsurl = {https://ui.adsabs.harvard.edu/abs/2003ApJ...591..288H}
}

@INPROCEEDINGS{hydra,
       author = {{Barden}, Samuel C.},
        title = "{The Use and Benefits of Optical Fibers in Spectroscopy}",
    booktitle = {Optical Astronomy from the Earth and Moon},
         year = 1994,
       editor = {{Pyper}, Diane M. and {Angione}, Ronald J.},
       series = {Astronomical Society of the Pacific Conference Series},
       volume = {55},
        month = jan,
        pages = {130-138},
       adsurl = {https://ui.adsabs.harvard.edu/abs/1994ASPC...55..130B}
}

@ARTICLE{kotak2006,
       author = {{Kotak}, R. and {Vink}, J.~S.},
        title = "{Luminous blue variables as the progenitors of supernovae with quasi-periodic radio modulations}",
      journal = {\aap},
         year = 2006,
        month = dec,
       volume = {460},
       number = {2},
        pages = {L5-L8},
          doi = {10.1051/0004-6361:20065800},
archivePrefix = {arXiv},
       eprint = {astro-ph/0610095},
 primaryClass = {astro-ph},
       adsurl = {https://ui.adsabs.harvard.edu/abs/2006A&A...460L...5K}
}

@ARTICLE{kuncarayakti2018,
       author = {{Kuncarayakti}, Hanindyo and {Maeda}, Keiichi and {Ashall}, Christopher J. and {Prentice}, Simon J. and {Mattila}, Seppo and {Kankare}, Erkki and {Fransson}, Claes and {Lundqvist}, Peter and {Pastorello}, Andrea and {Leloudas}, Giorgos and {Anderson}, Joseph P. and {Benetti}, Stefano and {Bersten}, Melina C. and {Cappellaro}, Enrico and {Cartier}, R{\'e}gis and {Denneau}, Larry and {Della Valle}, Massimo and {Elias-Rosa}, Nancy and {Folatelli}, Gast{\'o}n and {Fraser}, Morgan and {Galbany}, Llu{\'\i}s and {Gall}, Christa and {Gal-Yam}, Avishay and {Guti{\'e}rrez}, Claudia P. and {Hamanowicz}, Aleksandra and {Heinze}, Ari and {Inserra}, Cosimo and {Kangas}, Tuomas and {Mazzali}, Paolo and {Melandri}, Andrea and {Pignata}, Giuliano and {Rest}, Armin and {Reynolds}, Thomas and {Roy}, Rupak and {Smartt}, Stephen J. and {Smith}, Ken W. and {Sollerman}, Jesper and {Somero}, Auni and {Stalder}, Brian and {Stritzinger}, Maximilian and {Taddia}, Francesco and {Tomasella}, Lina and {Tonry}, John and {Weiland}, Henry and {Young}, David R.},
        title = "{SN 2017dio: A Type-Ic Supernova Exploding in a Hydrogen-rich Circumstellar Medium}",
      journal = {\apjl},
         year = 2018,
        month = feb,
       volume = {854},
       number = {1},
          eid = {L14},
        pages = {L14},
          doi = {10.3847/2041-8213/aaaa1a},
archivePrefix = {arXiv},
       eprint = {1712.00027},
 primaryClass = {astro-ph.SR},
       adsurl = {https://ui.adsabs.harvard.edu/abs/2018ApJ...854L..14K}
}

@ARTICLE{kuncarayakti2022,
       author = {{Kuncarayakti}, H. and {Maeda}, K. and {Dessart}, L. and {Nagao}, T. and {Fulton}, M. and {Guti{\'e}rrez}, C.~P. and {Huber}, M.~E. and {Young}, D.~R. and {Kotak}, R. and {Mattila}, S. and {Anderson}, J.~P. and {Ferrari}, L. and {Folatelli}, G. and {Gao}, H. and {Magnier}, E. and {Smith}, K.~W. and {Srivastav}, S.},
        title = "{Late-time H/He-poor Circumstellar Interaction in the Type Ic Supernova SN 2021ocs: An Exposed Oxygen-Magnesium Layer and Extreme Stripping of the Progenitor}",
      journal = {\apjl},
         year = 2022,
        month = dec,
       volume = {941},
       number = {2},
          eid = {L32},
        pages = {L32},
          doi = {10.3847/2041-8213/aca672},
archivePrefix = {arXiv},
       eprint = {2210.01755},
 primaryClass = {astro-ph.SR},
       adsurl = {https://ui.adsabs.harvard.edu/abs/2022ApJ...941L..32K}
}

@ARTICLE{kuncarayakti2023,
       author = {{Kuncarayakti}, H. and {Sollerman}, J. and {Izzo}, L. and {Maeda}, K. and {Yang}, S. and {Schulze}, S. and {Angus}, C.~R. and {Aubert}, M. and {Auchettl}, K. and {Della Valle}, M. and {Dessart}, L. and {Hinds}, K. and {Kankare}, E. and {Kawabata}, M. and {Lundqvist}, P. and {Nakaoka}, T. and {Perley}, D. and {Raimundo}, S.~I. and {Strotjohann}, N.~L. and {Taguchi}, K. and {Cai}, Y. -Z. and {Charalampopoulos}, P. and {Fang}, Q. and {Fraser}, M. and {Guti{\'e}rrez}, C.~P. and {Imazawa}, R. and {Kangas}, T. and {Kawabata}, K.~S. and {Kotak}, R. and {Kravtsov}, T. and {Matilainen}, K. and {Mattila}, S. and {Moran}, S. and {Murata}, I. and {Salmaso}, I. and {Anderson}, J.~P. and {Ashall}, C. and {Bellm}, E.~C. and {Benetti}, S. and {Chambers}, K.~C. and {Chen}, T. -W. and {Coughlin}, M. and {De Colle}, F. and {Fremling}, C. and {Galbany}, L. and {Gal-Yam}, A. and {Gromadzki}, M. and {Groom}, S.~L. and {Hajela}, A. and {Inserra}, C. and {Kasliwal}, M.~M. and {Mahabal}, A.~A. and {Martin-Carrillo}, A. and {Moore}, T. and {M{\"u}ller-Bravo}, T.~E. and {Nicholl}, M. and {Ragosta}, F. and {Riddle}, R.~L. and {Sharma}, Y. and {Srivastav}, S. and {Stritzinger}, M.~D. and {Wold}, A. and {Young}, D.~R.},
        title = "{The broad-lined Type-Ic supernova SN 2022xxf and its extraordinary two-humped light curves. I. Signatures of H/He-free interaction in the first four months}",
      journal = {\aap},
         year = 2023,
        month = oct,
       volume = {678},
          eid = {A209},
        pages = {A209},
          doi = {10.1051/0004-6361/202346526},
archivePrefix = {arXiv},
       eprint = {2303.16925},
 primaryClass = {astro-ph.SR},
       adsurl = {https://ui.adsabs.harvard.edu/abs/2023A&A...678A.209K}
}

@ARTICLE{kurita2020,
       author = {{Kurita}, Mikio and {Kino}, Masaru and {Iwamuro}, Fumihide and {Ohta}, Kouji and {Nogami}, Daisaku and {Izumiura}, Hideyuki and {Yoshida}, Michitoshi and {Matsubayashi}, Kazuya and {Kuroda}, Daisuke and {Nakatani}, Yoshikazu and {Yamamoto}, Kodai and {Tsutsui}, Hironori and {Iribe}, Masatsugu and {Jikuya}, Ichiro and {Ohtani}, Hiroshi and {Shibata}, Kazunari and {Takahashi}, Keisuke and {Tokoro}, Hitoshi and {Maihara}, Toshinori and {Nagata}, Tetsuya},
        title = "{The Seimei telescope project and technical developments}",
      journal = {\pasj},
         year = 2020,
        month = jun,
       volume = {72},
       number = {3},
          eid = {48},
        pages = {48},
          doi = {10.1093/pasj/psaa036},
       adsurl = {https://ui.adsabs.harvard.edu/abs/2020PASJ...72...48K}
}

@ARTICLE{lu2022,
       author = {{Lu}, J. and {Morrell}, N. and {Burns}, C. and {Hsiao}, E. and {Suntzeff}, N. and {Baron}, E. and {Shappee}, B. and {Aldoroty}, L. and {Anderson}, J. and {Ashall}, C. and {Bersten}, M. and {Brown}, P. and {Burrow}, A. and {Clochiatti}, A. and {Davis}, S. and {DerKacy}, J. and {Do}, A. and {Folatelli}, G. and {Buron}, F.~F. and {Galbany}, L. and {Hoeflich}, P. and {Holmbo}, S. and {Karamehmetoglu}, E. and {Krisciunas}, K. and {Kumar}, S. and {Mazzali}, P. and {Pessi}, P. and {Phillips}, M. and {Pignata}, G. and {Piro}, A.~L. and {Polin}, A. and {Shahbandeh}, M. and {Stangl}, S. and {Stritzinger}, M. and {Teffs}, J. and {Tonry}, J. and {Tucker}, M. and {Uddin}, S. and {Yang}, J.},
        title = "{POISE Transient Classification Report for 2022-03-18}",
      journal = {Transient Name Server Classification Report},
         year = 2022,
        month = mar,
       volume = {2022-733},
        pages = {1},
       adsurl = {https://ui.adsabs.harvard.edu/abs/2022TNSCR.733....1L}
}

@ARTICLE{maeda2013_10jl,
       author = {{Maeda}, K. and {Nozawa}, T. and {Sahu}, D.~K. and {Minowa}, Y. and {Motohara}, K. and {Ueno}, I. and {Folatelli}, G. and {Pyo}, T. -S. and {Kitagawa}, Y. and {Kawabata}, K.~S. and {Anupama}, G.~C. and {Kozasa}, T. and {Moriya}, T.~J. and {Yamanaka}, M. and {Nomoto}, K. and {Bersten}, M. and {Quimby}, R. and {Iye}, M.},
        title = "{Properties of Newly Formed Dust Grains in the Luminous Type IIn Supernova 2010jl}",
      journal = {\apj},
         year = 2013,
        month = oct,
       volume = {776},
       number = {1},
          eid = {5},
        pages = {5},
          doi = {10.1088/0004-637X/776/1/5},
archivePrefix = {arXiv},
       eprint = {1308.0406},
 primaryClass = {astro-ph.SR},
       adsurl = {https://ui.adsabs.harvard.edu/abs/2013ApJ...776....5M}
}

@ARTICLE{maeda2013_radio,
       author = {{Maeda}, Keiichi},
        title = "{Probing Shock Breakout and Progenitors of Stripped-envelope Supernovae through their Early Radio Emissions}",
      journal = {\apj},
         year = 2013,
        month = jan,
       volume = {762},
       number = {1},
          eid = {14},
        pages = {14},
          doi = {10.1088/0004-637X/762/1/14},
archivePrefix = {arXiv},
       eprint = {1209.1904},
 primaryClass = {astro-ph.HE},
       adsurl = {https://ui.adsabs.harvard.edu/abs/2013ApJ...762...14M}
}

@ARTICLE{maeda2022_Iax,
       author = {{Maeda}, Keiichi and {Kawabata}, Miho},
        title = "{Properties of Type Iax Supernova 2019muj in the Late Phase: Existence, Nature, and Origin of the Iron-rich Dense Core}",
      journal = {\apj},
         year = 2022,
        month = dec,
       volume = {941},
       number = {1},
          eid = {15},
        pages = {15},
          doi = {10.3847/1538-4357/ac9df2},
archivePrefix = {arXiv},
       eprint = {2210.14390},
 primaryClass = {astro-ph.HE},
       adsurl = {https://ui.adsabs.harvard.edu/abs/2022ApJ...941...15M}
}

@ARTICLE{matsubayashi2019,
       author = {{Matsubayashi}, Kazuya and {Ohta}, Kouji and {Iwamuro}, Fumihide and {Iwata}, Ikuru and {Kambe}, Eiji and {Tsutsui}, Hironori and {Izumiura}, Hideyuki and {Yoshida}, Michitoshi and {Hattori}, Takashi},
        title = "{KOOLS-IFU: Kyoto Okayama Optical Low-dispersion Spectrograph with optical-fiber Integral Field Unit}",
      journal = {\pasj},
         year = 2019,
        month = oct,
       volume = {71},
       number = {5},
          eid = {102},
        pages = {102},
          doi = {10.1093/pasj/psz087},
archivePrefix = {arXiv},
       eprint = {1905.05430},
 primaryClass = {astro-ph.IM},
       adsurl = {https://ui.adsabs.harvard.edu/abs/2019PASJ...71..102M}
}

@ARTICLE{margutti2017,
       author = {{Margutti}, Raffaella and {Kamble}, A. and {Milisavljevic}, D. and {Zapartas}, E. and {de Mink}, S.~E. and {Drout}, M. and {Chornock}, R. and {Risaliti}, G. and {Zauderer}, B.~A. and {Bietenholz}, M. and {Cantiello}, M. and {Chakraborti}, S. and {Chomiuk}, L. and {Fong}, W. and {Grefenstette}, B. and {Guidorzi}, C. and {Kirshner}, R. and {Parrent}, J.~T. and {Patnaude}, D. and {Soderberg}, A.~M. and {Gehrels}, N.~C. and {Harrison}, F.},
        title = "{Ejection of the Massive Hydrogen-rich Envelope Timed with the Collapse of the Stripped SN 2014C}",
      journal = {\apj},
         year = 2017,
        month = feb,
       volume = {835},
       number = {2},
          eid = {140},
        pages = {140},
          doi = {10.3847/1538-4357/835/2/140},
archivePrefix = {arXiv},
       eprint = {1601.06806},
 primaryClass = {astro-ph.HE},
       adsurl = {https://ui.adsabs.harvard.edu/abs/2017ApJ...835..140M}
}

@ARTICLE{moore2023,
       author = {{Moore}, T. and {Smartt}, S.~J. and {Nicholl}, M. and {Srivastav}, S. and {Stevance}, H.~F. and {Jess}, D.~B. and {Grant}, S.~D.~T. and {Fulton}, M.~D. and {Rhodes}, L. and {Sim}, S.~A. and {Hirai}, R. and {Podsiadlowski}, P. and {Anderson}, J.~P. and {Ashall}, C. and {Bate}, W. and {Fender}, R. and {Guti{\'e}rrez}, C.~P. and {Howell}, D.~A. and {Huber}, M.~E. and {Inserra}, C. and {Leloudas}, G. and {Monard}, L.~A.~G. and {M{\"u}ller-Bravo}, T.~E. and {Shappee}, B.~J. and {Smith}, K.~W. and {Terreran}, G. and {Tonry}, J. and {Tucker}, M.~A. and {Young}, D.~R. and {Aamer}, A. and {Chen}, T. -W. and {Ragosta}, F. and {Galbany}, L. and {Gromadzki}, M. and {Harvey}, L. and {Hoeflich}, P. and {McCully}, C. and {Newsome}, M. and {Gonzalez}, E.~P. and {Pellegrino}, C. and {Ramsden}, P. and {P{\'e}rez-Torres}, M. and {Ridley}, E.~J. and {Sheng}, X. and {Weston}, J.},
        title = "{SN 2022jli: A Type Ic Supernova with Periodic Modulation of Its Light Curve and an Unusually Long Rise}",
      journal = {\apjl},
         year = 2023,
        month = oct,
       volume = {956},
       number = {1},
          eid = {L31},
        pages = {L31},
          doi = {10.3847/2041-8213/acfc25},
archivePrefix = {arXiv},
       eprint = {2309.12750},
 primaryClass = {astro-ph.HE},
       adsurl = {https://ui.adsabs.harvard.edu/abs/2023ApJ...956L..31M}
}

@ARTICLE{nagao2023,
       author = {{Nagao}, T. and {Kuncarayakti}, H. and {Maeda}, K. and {Moore}, T. and {Pastorello}, A. and {Mattila}, S. and {Uno}, K. and {Smartt}, S.~J. and {Sim}, S.~A. and {Ferrari}, L. and {Tomasella}, L. and {Anderson}, J.~P. and {Chen}, T. -W. and {Galbany}, L. and {Gao}, H. and {Gromadzki}, M. and {Guti{\'e}rrez}, C.~P. and {Inserra}, C. and {Kankare}, E. and {Magnier}, E.~A. and {M{\"u}ller-Bravo}, T.~E. and {Reguitti}, A. and {Young}, D.~R.},
        title = "{Photometry and spectroscopy of the Type Icn supernova 2021ckj. The diverse properties of the ejecta and circumstellar matter of Type Icn supernovae}",
      journal = {\aap},
         year = 2023,
        month = may,
       volume = {673},
          eid = {A27},
        pages = {A27},
          doi = {10.1051/0004-6361/202346084},
archivePrefix = {arXiv},
       eprint = {2303.07721},
 primaryClass = {astro-ph.HE},
       adsurl = {https://ui.adsabs.harvard.edu/abs/2023A&A...673A..27N}
}

@ARTICLE{nagele2024,
       author = {{Nagele}, Chris and {Umeda}, Hideyuki and {Maeda}, Keiichi},
        title = "{STELLA Lightcurves of Energetic Pair-instability Supernovae in the Context of SN2018ibb}",
      journal = {\apj},
         year = 2024,
        month = sep,
       volume = {972},
       number = {1},
          eid = {11},
        pages = {11},
          doi = {10.3847/1538-4357/ad656c},
archivePrefix = {arXiv},
       eprint = {2404.16570},
 primaryClass = {astro-ph.HE},
       adsurl = {https://ui.adsabs.harvard.edu/abs/2024ApJ...972...11N}
}

@ARTICLE{ouchi2019,
       author = {{Ouchi}, Ryoma and {Maeda}, Keiichi},
        title = "{Constraining Massive Star Activities in the Final Years through Properties of Supernovae and Their Progenitors}",
      journal = {\apj},
         year = 2019,
        month = jun,
       volume = {877},
       number = {2},
          eid = {92},
        pages = {92},
          doi = {10.3847/1538-4357/ab1a37},
archivePrefix = {arXiv},
       eprint = {1904.07878},
 primaryClass = {astro-ph.HE},
       adsurl = {https://ui.adsabs.harvard.edu/abs/2019ApJ...877...92O}
}

@ARTICLE{pastorello2007,
       author = {{Pastorello}, A. and {Smartt}, S.~J. and {Mattila}, S. and {Eldridge}, J.~J. and {Young}, D. and {Itagaki}, K. and {Yamaoka}, H. and {Navasardyan}, H. and {Valenti}, S. and {Patat}, F. and {Agnoletto}, I. and {Augusteijn}, T. and {Benetti}, S. and {Cappellaro}, E. and {Boles}, T. and {Bonnet-Bidaud}, J. -M. and {Botticella}, M.~T. and {Bufano}, F. and {Cao}, C. and {Deng}, J. and {Dennefeld}, M. and {Elias-Rosa}, N. and {Harutyunyan}, A. and {Keenan}, F.~P. and {Iijima}, T. and {Lorenzi}, V. and {Mazzali}, P.~A. and {Meng}, X. and {Nakano}, S. and {Nielsen}, T.~B. and {Smoker}, J.~V. and {Stanishev}, V. and {Turatto}, M. and {Xu}, D. and {Zampieri}, L.},
        title = "{A giant outburst two years before the core-collapse of a massive star}",
      journal = {\nat},
         year = 2007,
        month = jun,
       volume = {447},
       number = {7146},
        pages = {829-832},
          doi = {10.1038/nature05825},
archivePrefix = {arXiv},
       eprint = {astro-ph/0703663},
 primaryClass = {astro-ph},
       adsurl = {https://ui.adsabs.harvard.edu/abs/2007Natur.447..829P}
}

@ARTICLE{pellegrino2022,
       author = {{Pellegrino}, C. and {Howell}, D.~A. and {Terreran}, G. and {Arcavi}, I. and {Bostroem}, K.~A. and {Brown}, P.~J. and {Burke}, J. and {Dong}, Y. and {Gilkis}, A. and {Hiramatsu}, D. and {Hosseinzadeh}, G. and {McCully}, C. and {Modjaz}, M. and {Newsome}, M. and {Gonzalez}, E. Padilla and {Pritchard}, T.~A. and {Sand}, D.~J. and {Valenti}, S. and {Williamson}, M.},
        title = "{The Diverse Properties of Type Icn Supernovae Point to Multiple Progenitor Channels}",
      journal = {\apj},
         year = 2022,
        month = oct,
       volume = {938},
       number = {1},
          eid = {73},
        pages = {73},
          doi = {10.3847/1538-4357/ac8ff6},
archivePrefix = {arXiv},
       eprint = {2205.07894},
 primaryClass = {astro-ph.HE},
       adsurl = {https://ui.adsabs.harvard.edu/abs/2022ApJ...938...73P}
}

@ARTICLE{perley2022,
       author = {{Perley}, Daniel A. and {Sollerman}, Jesper and {Schulze}, Steve and {Yao}, Yuhan and {Fremling}, Christoffer and {Gal-Yam}, Avishay and {Ho}, Anna Y.~Q. and {Yang}, Yi and {Kool}, Erik C. and {Irani}, Ido and {Yan}, Lin and {Andreoni}, Igor and {Baade}, Dietrich and {Bellm}, Eric C. and {Brink}, Thomas G. and {Chen}, Ting-Wan and {Cikota}, Aleksandar and {Coughlin}, Michael W. and {Dahiwale}, Aishwarya and {Dekany}, Richard and {Duev}, Dmitry A. and {Filippenko}, Alexei V. and {Hoeflich}, Peter and {Kasliwal}, Mansi M. and {Kulkarni}, S.~R. and {Lunnan}, Ragnhild and {Masci}, Frank J. and {Maund}, Justyn R. and {Medford}, Michael S. and {Riddle}, Reed and {Rosnet}, Philippe and {Shupe}, David L. and {Strotjohann}, Nora Linn and {Tzanidakis}, Anastasios and {Zheng}, WeiKang},
        title = "{The Type Icn SN 2021csp: Implications for the Origins of the Fastest Supernovae and the Fates of Wolf-Rayet Stars}",
      journal = {\apj},
         year = 2022,
        month = mar,
       volume = {927},
       number = {2},
          eid = {180},
        pages = {180},
          doi = {10.3847/1538-4357/ac478e},
archivePrefix = {arXiv},
       eprint = {2111.12110},
 primaryClass = {astro-ph.HE},
       adsurl = {https://ui.adsabs.harvard.edu/abs/2022ApJ...927..180P}
}

@ARTICLE{prestwich2007,
       author = {{Prestwich}, A.~H. and {Kilgard}, R. and {Crowther}, P.~A. and {Carpano}, S. and {Pollock}, A.~M.~T. and {Zezas}, A. and {Saar}, S.~H. and {Roberts}, T.~P. and {Ward}, M.~J.},
        title = "{The Orbital Period of the Wolf-Rayet Binary IC 10 X-1: Dynamic Evidence that the Compact Object Is a Black Hole}",
      journal = {\apjl},
         year = 2007,
        month = nov,
       volume = {669},
       number = {1},
        pages = {L21-L24},
          doi = {10.1086/523755},
archivePrefix = {arXiv},
       eprint = {0709.2892},
 primaryClass = {astro-ph},
       adsurl = {https://ui.adsabs.harvard.edu/abs/2007ApJ...669L..21P}
}

@ARTICLE{cartier2024,
       author = {{Cartier}, R{\'e}gis and {Contreras}, Carlos and {Stritzinger}, Maximilian and {Hamuy}, Mario and {Ruiz-Lapuente}, Pilar and {Prieto}, Jose L. and {Anderson}, Joseph P. and {Cikota}, Aleksandar and {Gerlach}, Matthias},
        title = "{Unveiling the nature of SN 2022jli: the first double-peaked stripped-envelope supernova showing periodic undulations and dust emission at late times}",
      journal = {arXiv e-prints},
         year = 2024,
        month = oct,
          eid = {arXiv:2410.21381},
        pages = {arXiv:2410.21381},
          doi = {10.48550/arXiv.2410.21381},
archivePrefix = {arXiv},
       eprint = {2410.21381},
 primaryClass = {astro-ph.HE},
       adsurl = {https://ui.adsabs.harvard.edu/abs/2024arXiv241021381C}
}

@ARTICLE{ryder2004,
       author = {{Ryder}, Stuart D. and {Sadler}, Elaine M. and {Subrahmanyan}, Ravi and {Weiler}, Kurt W. and {Panagia}, Nino and {Stockdale}, Christopher},
        title = "{Modulations in the radio light curve of the Type IIb supernova 2001ig: evidence for a Wolf-Rayet binary progenitor?}",
      journal = {\mnras},
         year = 2004,
        month = apr,
       volume = {349},
       number = {3},
        pages = {1093-1100},
          doi = {10.1111/j.1365-2966.2004.07589.x},
archivePrefix = {arXiv},
       eprint = {astro-ph/0401135},
 primaryClass = {astro-ph},
       adsurl = {https://ui.adsabs.harvard.edu/abs/2004MNRAS.349.1093R}
}

@ARTICLE{shingles2021,
       author = {{Shingles}, L. and {Smith}, K.~W. and {Young}, D.~R. and {Smartt}, S.~J. and {Tonry}, J. and {Denneau}, L. and {Heinze}, A. and {Weiland}, H. and {Flewelling}, H. and {Stalder}, B. and {Clocchiatti}, A. and {F{\"o}rster}, F. and {Pignata}, G. and {Rest}, A. and {Anderson}, J. and {Stubbs}, C. and {Erasmus}, N.},
        title = "{Release of the ATLAS Forced Photometry server for public use}",
      journal = {Transient Name Server AstroNote},
         year = 2021,
        month = jan,
       volume = {7},
        pages = {1-7},
       adsurl = {https://ui.adsabs.harvard.edu/abs/2021TNSAN...7....1S}
}

@ARTICLE{schulze2024,
       author = {{Schulze}, Steve and {Fransson}, Claes and {Kozyreva}, Alexandra and {Chen}, Ting-Wan and {Yaron}, Ofer and {Jerkstrand}, Anders and {Gal-Yam}, Avishay and {Sollerman}, Jesper and {Yan}, Lin and {Kangas}, Tuomas and {Leloudas}, Giorgos and {Omand}, Conor M.~B. and {Smartt}, Stephen J. and {Yang}, Yi and {Nicholl}, Matt and {Sarin}, Nikhil and {Yao}, Yuhan and {Brink}, Thomas G. and {Sharon}, Amir and {Rossi}, Andrea and {Chen}, Ping and {Chen}, Zhihao and {Cikota}, Aleksandar and {De}, Kishalay and {Drake}, Andrew J. and {Filippenko}, Alexei V. and {Fremling}, Christoffer and {Fr{\'e}our}, Laurane and {Fynbo}, Johan P.~U. and {Ho}, Anna Y.~Q. and {Inserra}, Cosimo and {Irani}, Ido and {Kuncarayakti}, Hanindyo and {Lunnan}, Ragnhild and {Mazzali}, Paolo and {Ofek}, Eran O. and {Palazzi}, Eliana and {Perley}, Daniel A. and {Pursiainen}, Miika and {Rothberg}, Barry and {Shingles}, Luke J. and {Smith}, Ken and {Taggart}, Kirsty and {Tartaglia}, Leonardo and {Zheng}, WeiKang and {Anderson}, Joseph P. and {Cassara}, Letizia and {Christensen}, Eric and {George Djorgovski}, S. and {Galbany}, Llu{\'\i}s and {Gkini}, Anamaria and {Graham}, Matthew J. and {Gromadzki}, Mariusz and {Groom}, Steven L. and {Hiramatsu}, Daichi and {Andrew Howell}, D. and {Kasliwal}, Mansi M. and {McCully}, Curtis and {M{\"u}ller-Bravo}, Tom{\'a}s E. and {Paiano}, Simona and {Paraskeva}, Emmanouela and {Pessi}, Priscila J. and {Polishook}, David and {Rau}, Arne and {Rigault}, Mickael and {Rusholme}, Ben},
        title = "{1100 days in the life of the supernova 2018ibb. The best pair-instability supernova candidate, to date}",
      journal = {\aap},
         year = 2024,
        month = mar,
       volume = {683},
          eid = {A223},
        pages = {A223},
          doi = {10.1051/0004-6361/202346855},
archivePrefix = {arXiv},
       eprint = {2305.05796},
 primaryClass = {astro-ph.HE},
       adsurl = {https://ui.adsabs.harvard.edu/abs/2024A&A...683A.223S}
}

@ARTICLE{schulze2025,
       author = {{Schulze}, Steve and {Gal-Yam}, Avishay and {Dessart}, Luc and {Miller}, Adam A. and {Woosley}, Stan E. and {Yang}, Yi and {Bulla}, Mattia and {Yaron}, Ofer and {Sollerman}, Jesper and {Filippenko}, Alexei V. and {Hinds}, K-Ryan and {Perley}, Daniel A. and {Tsuna}, Daichi and {Lunnan}, Ragnhild and {Sarin}, Nikhil and {Brennan}, Sean J. and {Brink}, Thomas G. and {Bruch}, Rachel J. and {Chen}, Ping and {Das}, Kaustav K. and {Dhawan}, Suhail and {Fransson}, Claes and {Fremling}, Christoffer and {Gangopadhyay}, Anjasha and {Irani}, Ido and {Jerkstrand}, Anders and {Knezevic}, Nikola and {Kushnir}, Doron and {Maeda}, Keiichi and {Maguire}, Kate and {Ofek}, Eran and {Omand}, Conor M.~B. and {Qin}, Yu-Jing and {Sharma}, Yashvi and {Sit}, Tawny and {Srinivasaragavan}, Gokul P. and {Strothjohann}, Nora L. and {Takei}, Yuki and {Waxman}, Eli and {Yan}, Lin and {Yao}, Yuhan and {Zheng}, WeiKang and {Zimmerman}, Erez A. and {Bellm}, Eric C. and {Coughlin}, Michael W. and {Masci}, Frank. J. and {Purdum}, Josiah and {Rigault}, Mickael and {Wold}, Avery and {Kulkarni}, Shrinivas R.},
        title = "{A cosmic formation site of silicon and sulphur revealed by a new type of supernova explosion}",
      journal = {arXiv e-prints},
         year = 2024,
        month = sep,
          eid = {arXiv:2409.02054},
        pages = {arXiv:2409.02054},
          doi = {10.48550/arXiv.2409.02054},
archivePrefix = {arXiv},
       eprint = {2409.02054},
 primaryClass = {astro-ph.HE},
       adsurl = {https://ui.adsabs.harvard.edu/abs/2024arXiv240902054S}
}

@ARTICLE{soderberg2006,
       author = {{Soderberg}, A.~M. and {Chevalier}, R.~A. and {Kulkarni}, S.~R. and {Frail}, D.~A.},
        title = "{The Radio and X-Ray Luminous SN 2003bg and the Circumstellar Density Variations around Radio Supernovae}",
      journal = {\apj},
         year = 2006,
        month = nov,
       volume = {651},
       number = {2},
        pages = {1005-1018},
          doi = {10.1086/507571},
archivePrefix = {arXiv},
       eprint = {astro-ph/0512413},
 primaryClass = {astro-ph},
       adsurl = {https://ui.adsabs.harvard.edu/abs/2006ApJ...651.1005S}
}
\bibliographystyle{aasjournal}

\end{document}